\documentclass{aa}  
\usepackage{graphicx}
\usepackage{txfonts}
\usepackage{natbib}
\bibpunct{(}{)}{;}{a}{}{,} % to follow the A&A style

\usepackage{color}

\definecolor{paolo}{rgb}{0.0, 0., 1.0}%{0.75, 0.25, 0.0}

\def \fexviii {Fe\,{\sc xviii}}

\def \fexix {Fe\,{\sc xix}}
\def \fexv {Fe\,{\sc xv}}
\def \feix {Fe\,{\sc ix}}
\begin{document}

   \title{Coronal energy release by MHD avalanches}

   \subtitle{III. Identification of a reconnection outflow from a nanoflare}

   \author{G. Cozzo\inst{1,2}
          \and
          P. Pagano \inst{2,3}
          \and
          F. Reale \inst{2,3}
          \and
          P. Testa \inst{1}
          \and
           A. Petralia \inst{2}
          \and
          J. Martinez-Sykora \inst{4,5,6,7}
          \and
          V. Hansteen \inst{5,6,7}
          \and
          B. De Pontieu \inst{4,5,6}  
          }

   \institute{
             Harvard–Smithsonian Center for Astrophysics, 60 Garden St., Cambridge, MA 02193, USA \\
        \email{gabriele.cozzo@cfa.harvard.edu}
            \and INAF-Osservatorio Astronomico di Palermo, Piazza del Parlamento 1, I-90134 Palermo, Italy
            \and 
            Dipartimento di Fisica \& Chimica, Università di Palermo, Piazza del Parlamento 1, I-90134 Palermo, Italy
            \and Lockheed Martin Solar \& Astrophysics Laboratory, 3251 Hanover St, Palo Alto, CA 94304, USA
            \and Rosseland Centre for Solar Physics, University of Oslo, P.O. Box 1029 Blindern, N-0315 Oslo, Norway
            \and Institute of Theoretical Astrophysics, University of Oslo, P.O. Box 1029 Blindern, N-0315 Oslo, Norway
            \and SETI Institute, 339 Bernardo Ave, Suite 200, Mountain View, CA, 94043, United States
            }

   \date{Received September 15, 1996; accepted March 16, 1997}

% \abstract{}{}{}{}{} 
% 5 {} token are mandatory
 
  \abstract
  {%\paolo{\sout{Kink-instability in a coronal magnetic flux tube might propagate in an avalanche, and generate a nanoflare storm. Related small-angle  reconnections can drive collimated outflow jets, named `nanojets'.}} 
  Outflows perpendicular to the guide field are believed to be a possible signature of magnetic reconnection in the solar corona and specifically a way to detect the occurrence of ubiquitous small-angle magnetic reconnection. }
  %The detection and analysis of such reconnection nanojets within the solar corona are emerging as a potentially valuable diagnostic tool to probe the enigmatic phenomenon of coronal heating.}
  {%\paolo{\sout{We address the formation, evolution and forward modelling of nanojets during an MHD avalanche.}
  The aim of this work is to identify possible diagnostic techniques of such outflows in hot coronal loops with the Atmospheric Image Assembly (AIA) on-board the Solar Dynamics Observatory and the forthcoming MUltislit Solar Explorer (MUSE), in a realistically dynamic coronal loop environment in which a magnetohydrodynamic (MHD) avalanche is occurring.}
  {%We conducted comprehensive full 3D magnetohydrodynamic (MHD) simulations to model the interaction between two magnetic strands forming a coronal loop, subjected to a global MHD instability. Our simulation comprehensively considers a stratified atmospheric structure, encompassing a high-beta chromosphere, a narrow transition region, and a rarefied magnetized corona. 
  We consider a 3D MHD model  of two  magnetic flux tubes, including a stratified, radiative and thermal-conducting atmosphere, twisted by footpoint rotation. The faster rotating flux tube becomes kink-unstable and soon involves the other one in the avalanche.
  The turbulent decay of this magnetic structure on a global scale leads to the formation, fragmentation, and dissipation of current sheets driving impulsive heating akin to a nanoflare storm. We captured a clear outflow from a reconnection episode soon after the initial avalanche and synthesized its emission as detectable with AIA and MUSE.}
  {The outflow has a maximum temperature around $8\,\mathrm{MK}$, a total energy of $10^{24}\,\mathrm{erg}$, a velocity of a few hundred km/s, and a duration of less than 1 min. We show the emission in the AIA 94 \AA\ channel (\fexviii\ line) and in the MUSE 108 \AA\ \fexix\ spectral line. }
  {This outflow shares many features with nanojets recently detected at lower temperatures. Its low emission measure makes, however, its detection difficult with AIA, but Doppler shifts can be measured with MUSE. Conditions become different in a later steady state phase when the flux tubes are filled with denser and relatively cooler plasma.}


   \maketitle
%
%-------------------------------------------------------------------

\section{Introduction}
%Coronal loops are closed magnetic structures that fill the lower solar corona \citep{vaiana1973identification, reale2014coronal}. Besides their tenuous density, they exhibit very high temperatures, ranging from one million to tens million kelvin degrees \citep{peter2014discovery, grotrian1939sonne, edlen1943deutung}.
The interplay between coronal magnetic field and photospheric motions might explain the starkly high temperature measured in the solar corona (\citealt{Alfven1947,parker1988nanoflares}, \citealt{gudiksen2005ab},\citealt{klimchuk2015key,testareale2023xray}). 
In particular, coronal heating might be the global result of discrete heating events occurring where magnetic braiding induces small scale, sub arcsec (the typical loops strands cross-sections are of $10$–$100\,\mathrm{km}$ according to \citealt{beveridge2003model, klimchuk2008highly, vekstein2009probing}) current sheets \citep[DC heating,][]{klimchuk2009coronal, viall2011patterns}.
These elemental and localized events have been referred to as `nanoflares' \citep{parker1988nanoflares}, and they release small amounts of energy ($\sim 10^{24}\,\mathrm{erg}$), which is promptly spread along the reconnected field lines via thermal conduction  \citep[and, at least in some cases, also by accelerated particles; see e.g.,][]{testa2014evidence, testa2020iris, cho2023statistical, wright2017microflare, glesener2020accelerated, cooper2021nustar}.

While small bursts have been observed in various wavelengths within the upper transition region or lower corona \citep[e.g.,][]{testa2013observing, testa2014evidence}, and high temperatures of $10\,\mathrm{MK}$ have been indirectly deduced from X-ray observations \citep[e.g.,][]{reale2011solar, testa2012hinode, ishikawa2017detection}, there has been no conclusive evidence of the widespread nanoflare activity as hypothesized by Parker.

%Recently, possible direct probes for nanoflare activity have been found to be the so-called `nanojets': reconnection outflow jets generated through a process resembling a slingshot effect during reconnection. They consist of the simultaneous ejection of plasma in opposite directions from the reconnection point, reaching Alfvénic speeds. They are considered a potential signature of reconnection-based nanoflares associated with the process of heating coronal loops. 
The unprecedented high spatial and temporal resolution observations of the solar atmosphere with the Interface Region Imaging Spectrograph (IRIS) \citep{de2014interface} and the Atmospheric Image Assembly (AIA) on-board the Solar Dynamics Observatory (SDO) \citep{pesnell2012solar, lemen2012atmospheric} enabled the discovery of fast and bursty `nanojets' \citep{antolin2021reconnection, sukarmadji2022observations}, which have been interpreted as direct evidence of coronal heating by magnetic reconnection in braided magnetic structures, and in particular, as outflow jets accelerated by the slingshot effect of magnetic field lines during small-angle reconnection. Such episodic phenomena provide novel and important diagnostics of nanoflare activity, overcoming the general difficulties in directly observing nanoflares due to several factors, such as the efficient thermal conduction which rapidly washes out the evidence of the high temperature bursts produced by impulsive heating.

High resolution observations of active regions (\citealt{antolin2021reconnection}, \citealt{sukarmadji2022observations}, \citealt{patel2022hi}, \citealt{sukarmadji2024transverse}) have revealed a variety of small ($500-1500\,\mathrm{km}$), and transient ($< 30\,\mathrm{s}$) nanoflare-like EUV bursts followed by collimated outflows, named nanojets, 100 to 300 km/s fast, presumably driven by dynamic instabilities such as magnetohydrodynamic (MHD) avalanches \citep{antolin2021reconnection}, Kelvin-Helmholtz, and Rayleigh–Taylor instabilities \citep{sukarmadji2022observations} or during catastrophic cooling of coronal loop strands \citep{sukarmadji2024transverse}, often accompanied by the formation of coronal rain \citep{antolin2015multi}.
Observation of nanojets in different temperature channels supports the hypothesis of multi-thermal structuring \citep[e.g.,][]{sukarmadji2022observations, patel2022hi}, predominantly at temperatures around and below $1\,\mathrm{MK}$.
Although bidirectional jets are expected from reconnection, observed nanojets are often strongly asymmetric \citep[e.g.,][]{patel2022hi}, possibly due to loop’s curvature \citep{pagano2021modelling} or braiding.

%\cite{antolin2021reconnection} reports the observation of a multitude of nanojets on a loop structure monitored at the limb of the Sun on April 2014. Such observations were carried out with the Atmospheric Imaging Assembly (AIA, \cite{lemen2012atmospheric}) on board of the Solar Dynamics Observatory (SDO, \cite{pesnell2012solar}), the Interface Region Imaging Spectrograph (IRIS, \cite{de2014interface}) and the Hinode/Solar Optical Telescope (SOT, \cite{suematsu2008solar, tsuneta2008solar}). The detected sequence of events ultimately resulted in the creation of a highly heated coronal loop. The way in which these heating events evolved in space and time, along with the dynamics of the interwoven loop structure, exhibited features that align with the characteristics typically associated with a MHD avalanche.  Specifically, the observed nanojets exhibit specific characteristics, including confinement with dimensions typically around 500 km in width and lengths ranging from 1000 to 2000 km. They are short-lived, lasting approximately 15 seconds or less, and are characterized by rapid plasma flows of 100 to 200 km/s that move perpendicular to the magnetic field guiding the coronal loop \citep{antolin2021reconnection, sukarmadji2022observations}. 
Properties of such reconnecting plasma outflows were investigated via MHD numerical simulations  \citep[e.g.,][]{antolin2021reconnection, pagano2021modelling, de2022probing}. 
\cite{antolin2021reconnection} show a non-ideal MHD simulation of two interacting, gravitationally stratified coronal loops, the footpoints of which are slowly moved in opposite directions to create a small angle between the loops.  
As the x-type misalignment increases, the electric current between the loops increases as well, thus leading to magnetic field lines reconnection at the mid-plane. The enhanced magnetic tension in the reconnection region drives a transverse displacement of the plasma. A high-velocity (up to $200\,\mathrm{km}\,\mathrm{s}^{-1}$), collimated (widths of order of few $\mathrm{Mm}$), bidirectional jet also results from the reconnection process. 

The forthcoming MUltislit Solar Explorer \citep[MUSE,][]{de2019multi} will be able to provide key diagnostics of reconnection outflows, as shown in e.g. \cite{de2022probing}. 
%They predict MUSE synthetic observables from the 3D MHD model of \cite{antolin2021reconnection}. 
Distinctive signatures of the ongoing outflow appear in Doppler shifts and non-thermal line widths (e.g., small Doppler velocity and enhanced non-thermal line broadening at the reconnection site, respectively).
%However, it remains to test in a more dynamic and evolving environment whether it still possible to observationally single out nanojets.

\begin{figure}[h!]
   \centering
  \includegraphics[width=\hsize]{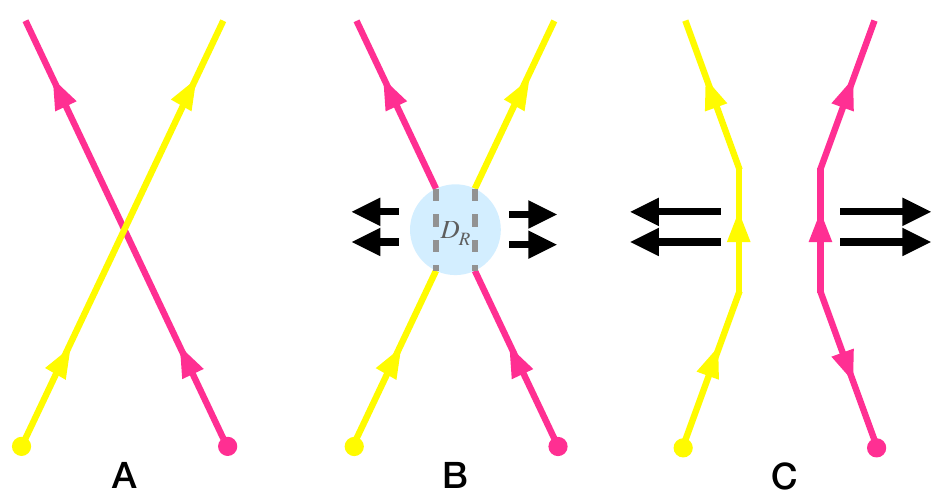}
  \caption{Schematic representation of guide field small-angle reconnection at three different stages. Step A: two field lines are tilted in opposite directions. Step B: field lines reconnect in the diffusion region $D_R$ where currents are stronger. Step C: after the reconnection the field line connectivity has changed. The black arrows indicate the outflows.}  
  \label{Fig:reconnection}
\end{figure}
A schematic description of the reconnection processes yielding the outflow acceleration is shown in Fig. \ref{Fig:reconnection}. As drifting magnetic field lines are driven towards one another (A), the magnetic field component perpendicular to the guide field vanishes at the `dissipation region'  \citep[$D_{\mathrm{R}}$, where $\vec E \cdot \vec B \ne 0$,][]{hesse1988theoretical, schindler1988general}, and it induces the field connectivity to change (B).
The magnetic field reconnects inside $D_{\mathrm{R}}$.
The new field lines configuration induces a magnetic tension imbalance which in turn drives field lines to expand outwards (C). Outside of the dissipation region, the plasma is frozen in the field and is accelerated by the slingshot effect caused by the released magnetic tension. 

Beyond the work already done in basic models, it remains to be addressed if reconnection collimated outflows can occur in more realistic and dynamic scenarios and whether even in these circumstances they can be detected and observed with current or upcoming instruments.
To answer these questions, this work addresses the MHD and forward modelling of an outflow, forming and evolving during an MHD Avalanche \citep{2016mhhoodd}. We investigate the full 3D MHD simulation of an MHD avalanche described in \cite{cozzo2023coronal} and check the occurrence of nanojets-like events during the evolution of instability. 
%In particular, in Sec. \ref{Sec:modelinganddiagnostics} we briefly describe the reference MHD simulation (we refer to \citealt{cozzo2023coronal} for more details). Then, we characterize the evolution of a nanojet evolving during the dynamic phase of an MHD-avalanche and show the synthetic observables with AIA and MUSE instruments. In Sec. \ref{Sec:Conclusions} we discuss the results and draw the conclusions.

\section{MHD modelling}
\label{Sec:modelinganddiagnostics}

%Coronal (loops) heating through reconnection-induced nanoflare storms can be significant only when the magnetic field has stored enough free energy to heat the coronal loop. A possible triggering mechanism is the `MHD Avalanche' model \citep{2016mhhoodd}: when the system attains a critical state in which the local loss of equilibrium within an elemental loop structure (strand) spreads throughout the entire structure, it leads to the onset of a nanoflare storm, ultimately heating the entire loop.

We consider the 3D MHD simulation
%of MHD avalanches involved strongly twisted, kink-unstable coronal loops, either single \citep{hood1979kink, hood2009coronal} or multi-threaded \citep{tam2015coronal, 2016mhhoodd, reid2018coronal, reid2020coronal}, dynamically decaying in a turbulent cascade of current sheets. Specifically, when multiple magnetic strands coexisted within the same coronal loop, a single unstable magnetic filament could disrupt nearby stable loops, initiating a chain reaction of global magnetic dissipation.
%These experiments focused solely on the coronal loops, without considering their interaction with the underlying chromospheric layer.
%More recently, \cite{cozzo2023coronal} carried out a MHD simulation of a stratified solar atmosphere encompassing chromospheric layer, transition region, and lower corona where multiple interacting coronal loop strands experience twisting and subsequently become unstable. Their findings provided confirmation that avalanches represent a viable mechanism for storing and releasing magnetic energy in coronal loops, as a consequence of photospheric motions.
%The model 
described in \cite{cozzo2023coronal} of a MHD avalanche in a kink unstable, multi-threaded coronal loop system \citep[see also,][]{2016mhhoodd, reid2018coronal, reid2020coronal}. 
In this case, small angle reconnection episodes result from the turbulent dissipation of the twisted magnetic field during the instability, rather than by regular photospheric motions, directly tilting the field lines, as in \cite{pagano2021modelling}, and \cite{de2022probing}.
Two identical magnetic flux tubes are 
%It addresses a multi-threaded coronal loop structure 
embedded in a stratified solar atmosphere with a $1\,\mathrm{MK}$ corona anchored on both sides to a  
%a million kelvin, magnetized corona and bounded by two 
dense and cooler isothermal ($10^4\,\mathrm{K}$) chromosphere. The flux tubes are progressively twisted at different angular velocities by photospheric rotation motions at their footpoints, mirrored with respect to the middle plane \citep{reale20163d, cozzo2023asymmetric}, in a background magnetic field ($B_{\mathrm{bkg}} = 10\,\mathrm{Gauss}$).
%We reproduced the dynamics of an avalanche-like event by numerical integration the non-ideal MHD equations (see eq.s (1)-(7) in \cite{cozzo2023coronal}) in a full 3D space and with Pluto code \citep{mignone2007pluto}. 
The time-dependent 3D MHD equations are solved with the Pluto code \citep{mignone2007pluto}.
The computational box has a size $\Delta x = 16\,\mathrm{Mm}$, $\Delta y = 8 \,\mathrm{Mm}$, $\Delta z = 62\,\mathrm{Mm}$.
%We assumed high length-width aspect ratio to keep intrinsic curvature effects negligible. 
%Radiative losses and thermal conduction drain internal energy from the corona while a uniform and static heating prevents this layer to cool down to chromospheric temperatures \citep{guarrasi2014mhd, cozzo2023asymmetric}. 
Anomalous magnetic resistivity \citep[$\eta_0 = 10^{14}\,\mathrm{cm}^{-2}\,\mathrm{s}^{-1}$,][]{hood2009coronal, reale20163d} turns on in the corona ($T > 10^4\,\mathrm{K}$) when and where the electric current  density $\vec j$ exceeds the threshold value $j_{\mathrm{cr}} =250\,\mathrm{Fr}\,\mathrm{cm}^{-2}\,\mathrm{s}^{-1}$.
The equations include radiative losses, thermal conduction, and gravity component for a curved flux tube (closed coronal loop). 
 % anomalous magnetic resistivity are taken into account, as in \cite{reale20163d} and \cite{cozzo2023asymmetric}.
%Two parallel magnetic flux tubes lie on a background magnetic field ($B_{bkg} = 10\,\mathrm{Gauss}$) one nearby each other, line-tied to the upper and lower boundaries of the computational box.
Due to progressive twisting, the faster rotating flux tube becomes kink-unstable and rapidly fragments into a chaotic system with thin current sheets hosting small-size impulsive reconnection events. The instability soon propagates to the nearby slower tube, which then evolves in a similar way.
%Circular photospheric motions induce twisting to each loop's strand. At a certain point, one magnetic strand overcomes the critical threshold for kink modes instability. As the twisted filament undergoes kink instability, the second flux tube is perturbed and rapidly departs from equilibrium.
%The following global MHD instability results into current sheets formation and reconnection of magnetic field lines.
%The dynamic dissipation of the multi-threaded coronal loop into smaller current sheets involves a series of irregular and discreet heating events.
The impulsive events cause local heating of the plasma to temperature peaks above $10\,\mathrm{MK}$.
%widespread occurrence of these intense bursts makes coronal temperature to increase
%and chromospheric plasma to evaporate, making in turns the loop brighter in the EUV band.

\begin{figure*}[h!]
   \centering
\includegraphics[width=\hsize]{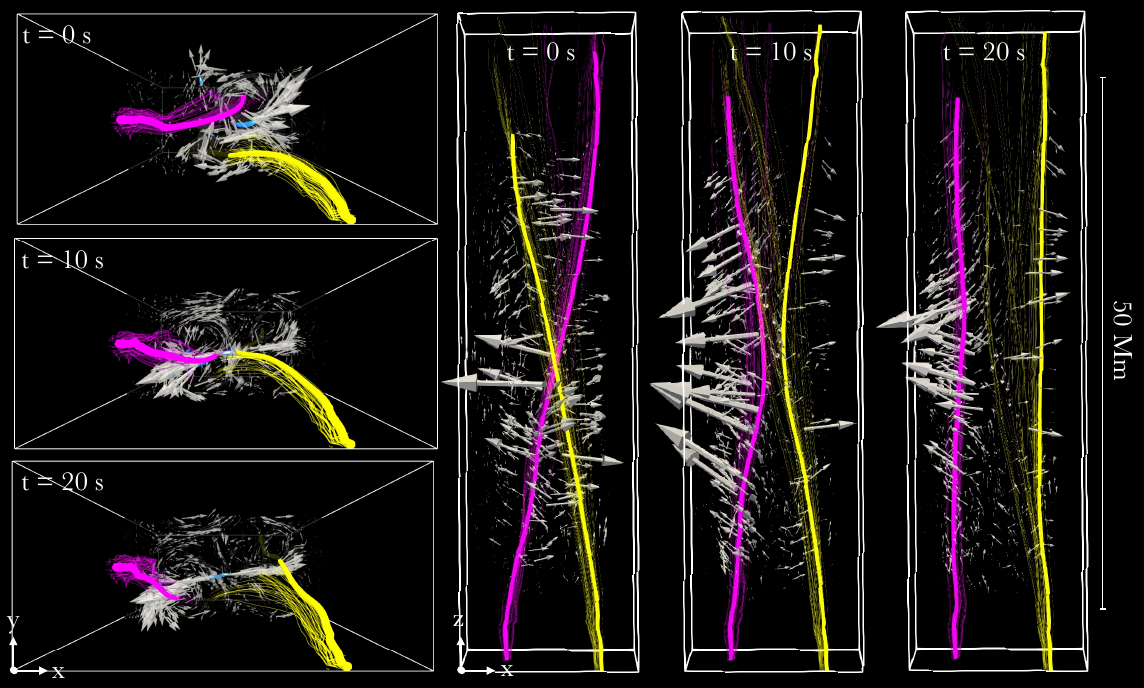}
  \caption{Magnetic reconnection and the outflow. Rendering 3D at 3 times since the beginning of the avalanche: $\Delta t = 0\,\mathrm{s}$ (lines approaching), $\Delta t = 10\,\mathrm{s}$ (lines reconnecting), and $\Delta t = 20\,\mathrm{s}$ (new lines detaching). Left column, top view and cut at the middle plane: two reconnecting magnetic field lines  (marked by yellow and magenta lines among two bundles), reconnection sites (blue spots close to the centre of the plane), and velocity field (white arrows), which shows the collimated outflow departing from the reconnecting lines (see Movie 1). Right row: same reconnecting lines from a front view (the coronal part of the loop is $50\,\mathrm{Mm}$ long, see Movie 2.)} 
  \label{Fig:Nanojets_3D}
\end{figure*}

\begin{figure*}[h!]
   \centering
  \includegraphics[width=\hsize]{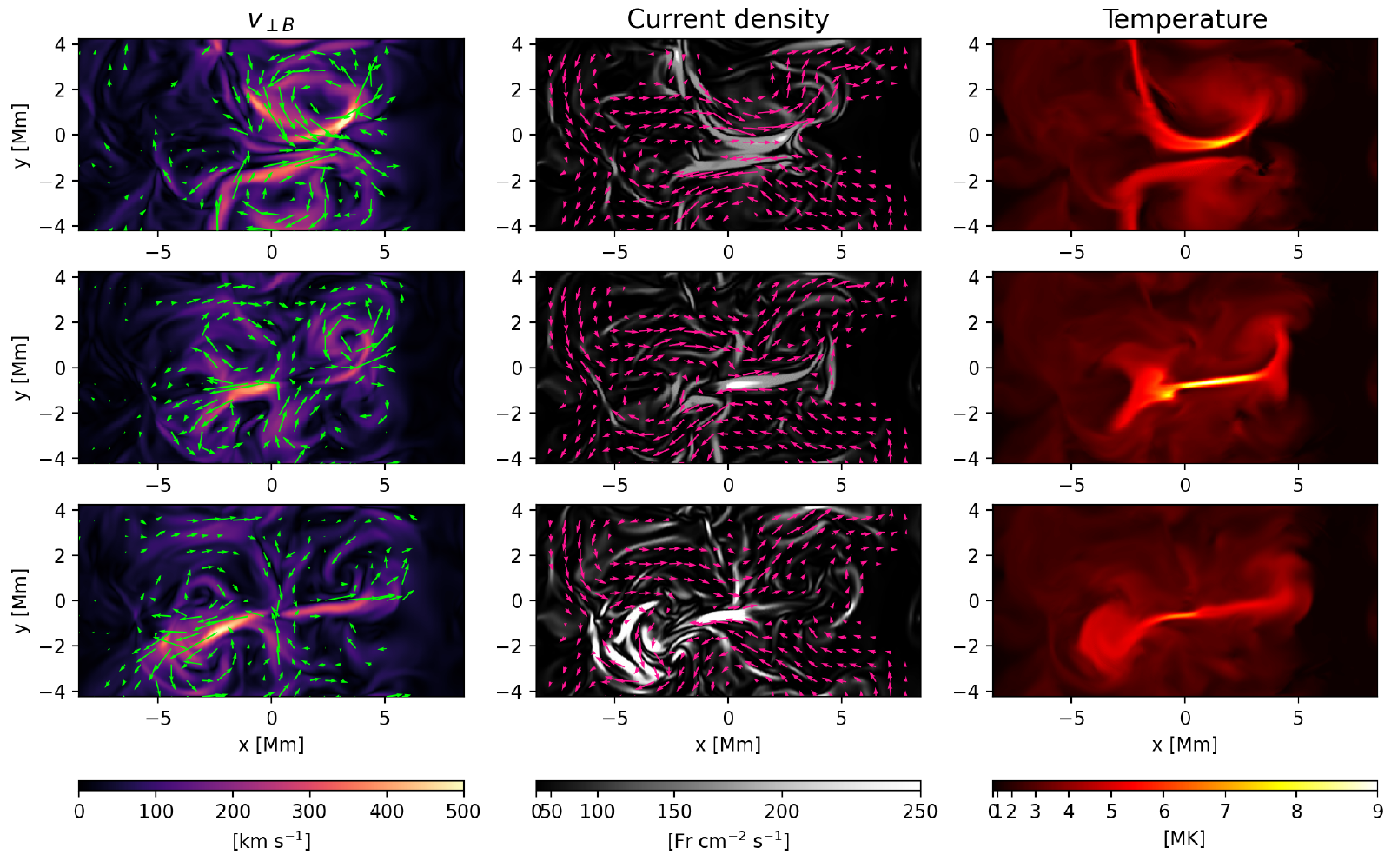}
  \caption{Dynamics of the reconnection outflow. First column: horizontal mid-plane map of the value of the velocity component perpendicular to the magnetic field at the same three times shown in Fig. \ref{Fig:Nanojets_3D}, i.e.,  $\Delta t = 0\,\mathrm{s}$ (top, current sheet formation), $\Delta t = 10\,\mathrm{s}$ (middle, outflow acceleration), and $\Delta t = 20\,\mathrm{s}$ (bottom, outflow deceleration). he velocity field in the plane is also shown (arrows). Second column: horizontal mid-plane map of the current density. The map saturates where the current density exceeds the threshold value for dissipation. The magnetic field in the plane is also shown (arrows). Third column: horizontal mid-plane map of the temperature. (See Movie 3.)}
  \label{Fig:dynamics_nanojet}
\end{figure*}

\begin{figure}[h!]
   \centering
  \includegraphics[width=\hsize]{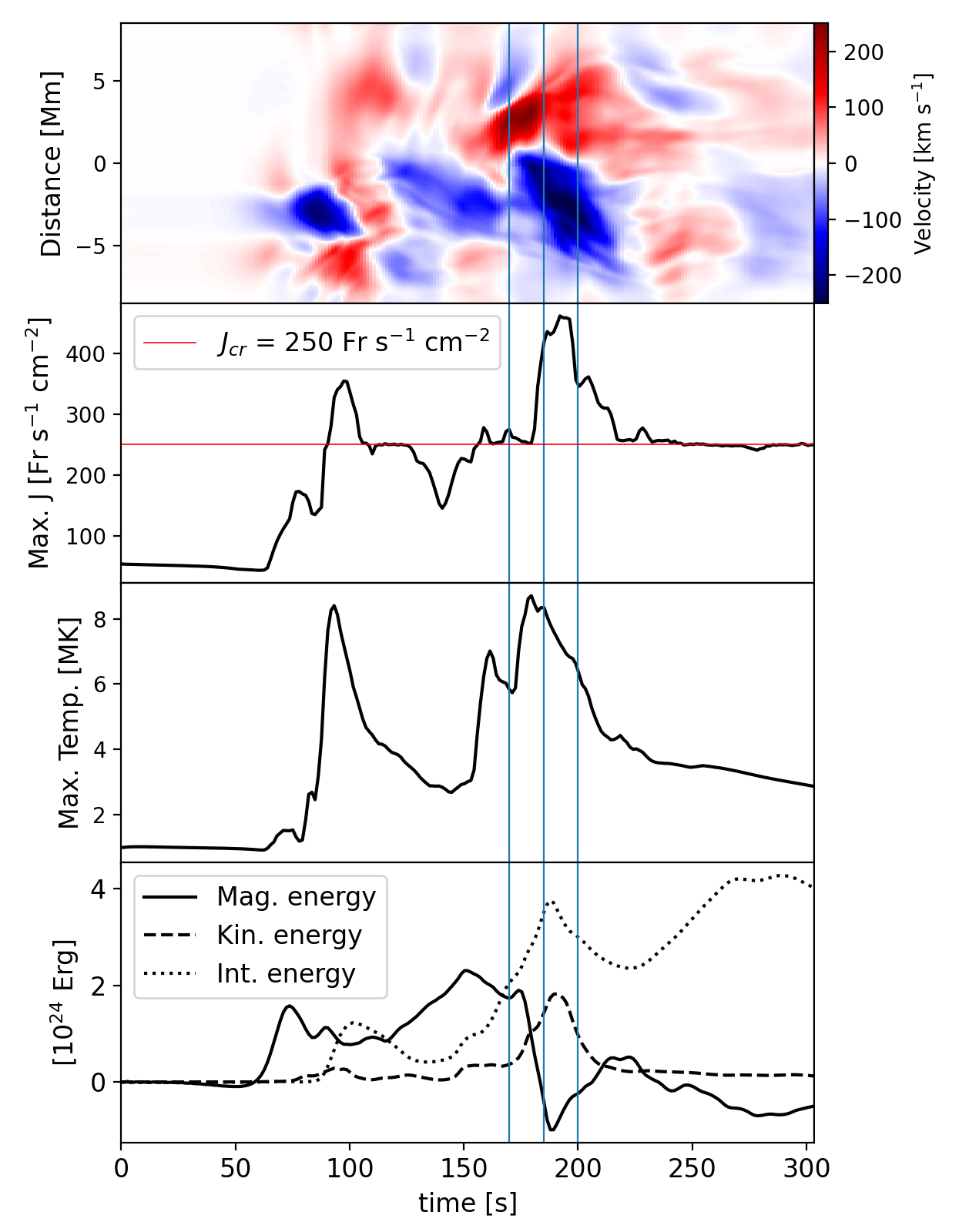}
  \caption{Evolution of the coronal loop plasma in a box containing the reconnection outflow. This reference volume has size $\Delta x = 10\,\mathrm{Mm}$, $\Delta y = 4 \,\mathrm{Mm}$, $\Delta z = 10\,\mathrm{Mm}$, and is centred at the origin. First panel: x-component of the velocity, averaged along $\Delta y$ and $\Delta z$. Second panel:  maximum temperature in the box. Third panel: maximum current density in the box. Fourth panel:
  total magnetic (solid), kinetic (dashed), and internal (dotted line) energy in the box.}
  \label{Fig:nanojet_time_evolution}
\end{figure}

%\begin{figure}[h!]
%   \centering
%  \includegraphics[width=\hsize]{Images/1D_plot_nanojet.png}
%  \caption{Evolution of the coronal loop inside a reference volume of size $\Delta x = 1.6 \times 10^9\,\mathrm{cm}$, $\Delta y = 0.5 \times 10^9\,\mathrm{cm}$, $\Delta z = 0.5 \times 10^9\,\mathrm{cm}$ centred at the origin of the computational box. From top to bottom: total magnetic (solid), kinetic (dashed), and internal energy (dotted) changes as function of time; maximum current density against time; maximum temperature against time. 
%  Blue, vertical lines show the time snapshot times used in Fig. \ref{Fig:Nanojets_3D} and \ref{Fig:dynamics_nanojet}. The red, horizontal line in the second panel indicates the threshold value for dissipation.}
%  \label{Fig:1D_plot_nanojet}
%\end{figure}

%\begin{figure}[h!]
%   \centering
%  \includegraphics[width=\hsize]{Images/Integrated_1_2r.pdf}
%  \caption{Evolution of the coronal loop inside a reference volume of size $\Delta x = 1.6 \times 10^9\,\mathrm{cm}$, $\Delta y = 0.5 \times 10^9\,\mathrm{cm}$, $\Delta z = 0.5 \times 10^9\,\mathrm{cm}$ centred at the origin of the computational box. From top to bottom: total magnetic, kinetic, and internal energy changes as function of time; 
  %maximum and averaged current density against time; maximum and averaged temperature as function of time. Blue, vertical lines encapsulate the time lapse when the nanojet is accelerated. 
  %The red, dashed, horizontal line in the second panel indicates the threshold value for dissipation.
%  }
%  \label{Fig:Integrated_1}
%\end{figure}

\begin{figure*}[h!]
   \centering
  \includegraphics[width=0.21\hsize]{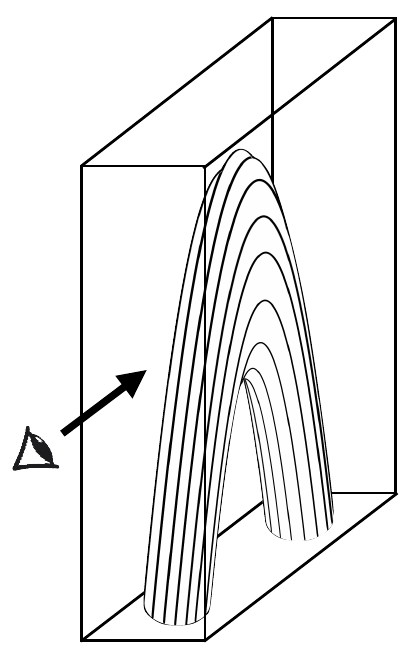}
  \includegraphics[width=0.78\hsize]{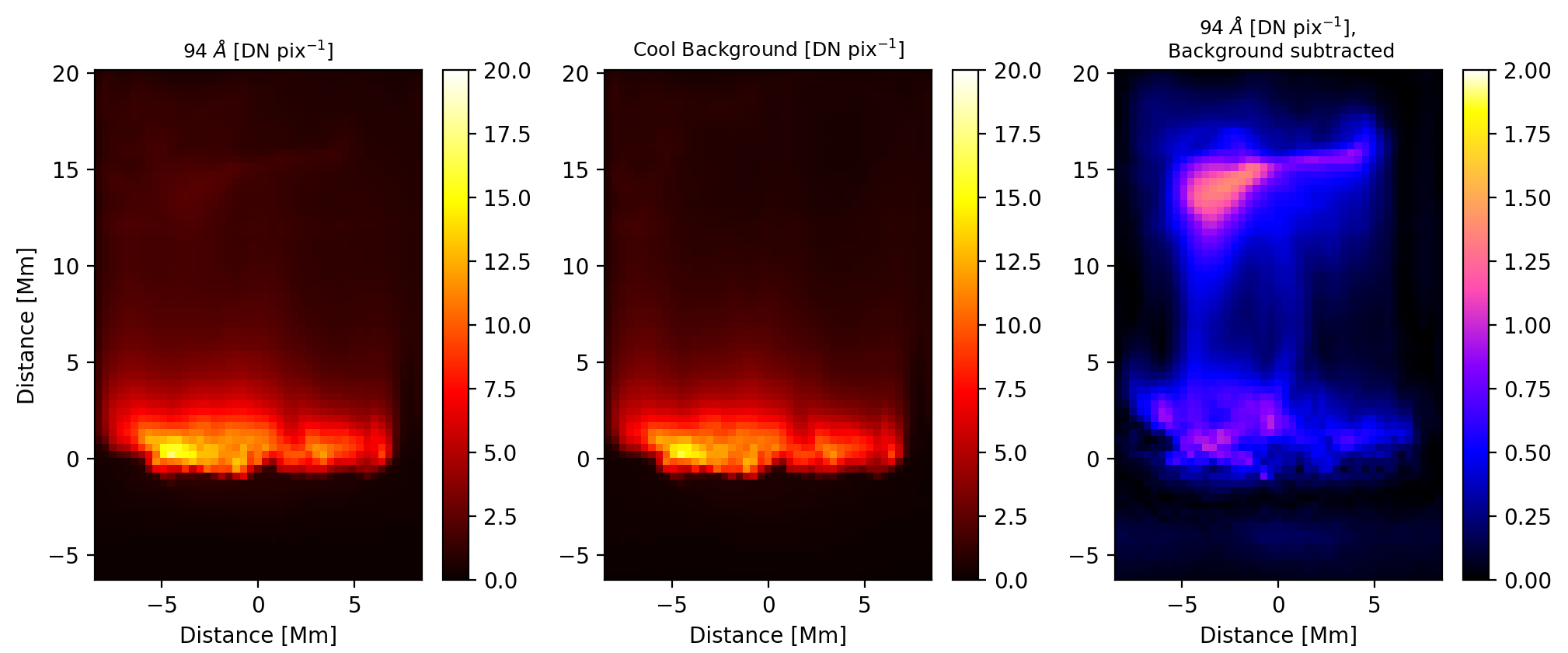}
  \caption{Synthetic maps in the AIA 94 \AA\ channel integrated along a line of sight from a side view of the curved loop system (first panel, exposure time: $9\,\mathrm{s}$). Second panel: In this geometry the top of the loop is high in the image, as shown in the left panel. intensity map in the entire filter band. Third panel: intensity map of the cool component ($\sim 1\,\mathrm{MK}$). Fourth panel: map of the hot ( \fexviii\ ) component only, after subtracting the middle from the left.}
  \label{Fig:AIA_diagnostics}
\end{figure*}

\begin{figure*}[h!]
   \centering
  \includegraphics[width=\hsize]{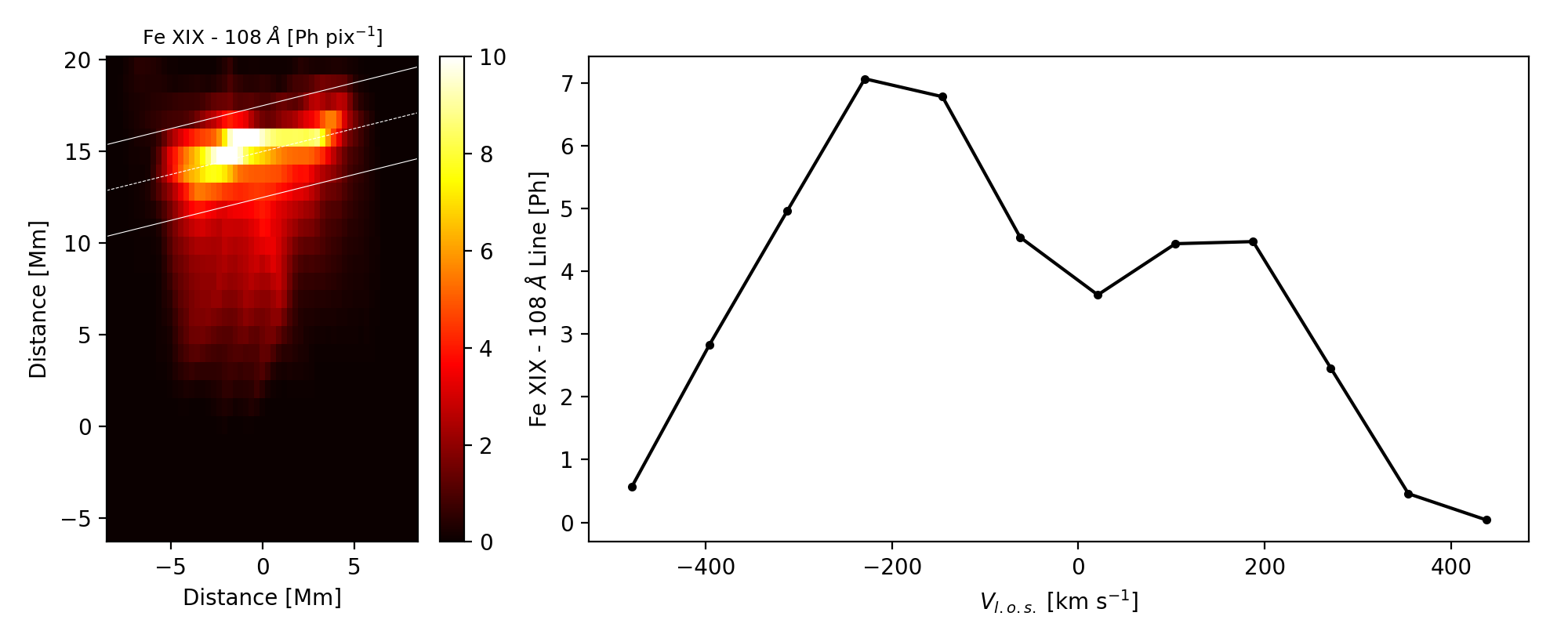}
  \caption{MUSE synthetic map and spectrum (line of sight shown in the fig. \ref{Fig:AIA_diagnostics}). Left:  the \fexix\ 108 \AA\ line emission map as in Fig.\ref{Fig:AIA_diagnostics}. The emission is integrated over macro-pixels of size $\Delta_h = 0.28$ Mm, $\Delta_v = 1.89$ Mm ($0.4" \times 2.7"$) (exposure time: $30\,\mathrm{s}$). Right: \fexix\ line spectrum obtained by integrating the emission along the volume marked in the map on the left (white solid lines, cross section $5\times5\,\mathrm{Mm}$  The spectral bin is $\Delta v = 80\,\mathrm{km}\,\mathrm{s}^{-1}$. We account for thermal, non-thermal and instrumental broadening.}
  \label{Fig:MUSE_diagnostics}
\end{figure*}

%\section{Nanojets}
%\label{Sec:Nanojetdynamics}
%In this paper we focus on the evolution of the MHD avalanche after the initial, violent decay phase of the twisted magnetic field structure. In particular, 
Following the evolution of several magnetic field lines in the aftermath of the MHD avalanche, we identified a few examples of reconnection jets, i.e. bundles of magnetic field lines whose evolution follows the patterns described in Fig.\ref{Fig:reconnection}.
We selected one of them as a reference caseesulting from the reconnection of two slightly misaligned magnetic filaments carried out by swirling plasma flows. 
%In particular, as the two magnetic flux tubes become closer, the current density rapidly grows up in the intermediate region as a results of the rapid settling down of a magnetic field tangential discontinuity. Consequently, a current sheet forms around the field lines contact point. As the shift between the flux ropes diminishes, the electric current within them proportionally escalates, ultimately exceeding the critical threshold for anomalous resistivity. Subsequently, magnetic field lines forming the two strands undergo reconnection, altering their trajectory nearby the mid-plane of the box. The modified topology releases magnetic tension in that zone, which in turns accelerates a transverse displacement of the plasma.

%Fig. \ref{Fig:Nanojets_3D} field lines in the box are shown in full-3D rendering. The field lines were computed using a fourth order Runge-Kutta scheme, and  colour was attributed depending on where on the photosphere the starting points were placed.

Figure \ref{Fig:Nanojets_3D} shows a localized reconnection event with the reference outflow. In the box ($62\,\mathrm{Mm}$ long, and 
$8\,\mathrm{Mm} \times 16\,\mathrm{Mm}$ cross section), the atmosphere is stratified, with two chromospheric layers at the top and bottom of the box, and a $50\,\mathrm{Mm}$ corona (see Appendix \ref{Sec:Athmospheric_stratification} for details). Field lines in the box are shown in full-3D rendering at three different times and from two different perspectives. They are computed using a fourth-order Runge-Kutta scheme, while the colour is attributed depending on the starting points at the lower photospheric boundary, which move following photospheric twisting. On the left, a view from the top, along the coronal loop axis (z-axis), with a cut at the mid-plane of the box showing in blue the electric field component parallel to the magnetic field. Arrows mark the orientation and strength of the velocity field. On the right, a front view.
We draw two reconnecting magnetic field lines. %\paolo{\sout{, computed with Runge-Rutta scheme of order 4 with starting points at the lower boundary of the domain}}.
%, where the plasma is stack into the deepest chromospheric layer.
The sites of reconnection are localized as those where the electric field component parallel to the magnetic field is non-zero  \citep[$E_{\parallel}$, blue spots in the left panels,][see also Fig. \ref{Fig:Nanojets_3D_A} for a full-3D rendering of $D_R$]{hesse1988theoretical, schindler1988general, reale20163d}. 
%White arrows show the orientation of the velocity field. 
The velocity field (arrows) illustrates the approaching flows and the collimated outflows diverging from the reconnection site. Movie 1 shows the evolution of the 3D rendering.
Initially, the two field lines approach each other, dragged by the chaotic dynamics of the MHD avalanche. The outflowing plasma is then accelerated (second panel) in the dissipation region near the mid-plane centre, i.e., where $E_{\parallel}$ becomes stronger. Afterwards (third panel) the reconnecting field pushes the plasma outwards where it eventually disperses in the ambient magnetic field.  

On the right (and in Movie 2), the front view shows the reconnecting lines emphasizing the presence of the guide field, similar to Fig. \ref{Fig:reconnection}. 
In the first panel, the reconnecting magnetic field is starting to push the plasma outwards (as emphasized by arrows in a few high velocity spots near the reconnection site). 
%The outflow develops from the strongest dynamics of the magnetic field, advecting plasma from the reconnection site, with small field-aligned plasma velocity.
The plasma velocity is mostly perpendicular to the field lines, with a small component along them.
This transverse motion is stronger (middle panel) and longer-lasting (third panel) around the middle of the flux tube.
In Fig. \ref{Fig:dynamics_nanojet} we illustrate details of the outflow from the selected reconnection event. The three columns show the velocity component perpendicular to the magnetic field ($v_{\perp}$), current density and temperature maps, on the mid-plane perpendicular to the $z$ vertical axis and at the same times as in Fig. \ref{Fig:Nanojets_3D}. Movie 3 shows the evolution of the same quantities.
%we illustrate the evolution of plasma and electric current during the magnetic field lines interaction that we have described.
%enters deeper into the nanojet analysis.
%In the first column the colour map is the value of the velocity component perpendicular to the magnetic field ($v_{\perp}$), and the arrows the velocity field (projected on the plane).
%are presented at the same times as in figure. \ref{Fig:Nanojets_3D}. 
%The  arrows show instead the orientation of the \paolo{whole velocity field vector}. [BOH?] 
The velocity maps $v_{\perp}$ emphasize where the plasma is accelerated by the magnetic field tension (as the Lorentz force acts always perpendicular to $\mathbf{B}$) and where instead the plasma drags the magnetic field lines. As expected, the velocity field diverges from the central reconnection site, near $x=0\,\mathrm{Mm}$ and $y=-1\,\mathrm{Mm}$ in two strong, sub-Alfvénic collimated jets. 
%The second column  of Fig. \ref{Fig:dynamics_nanojet} displays mid-plane current density, together with the magnetic field vectors. 
The current maps show an intense sheet in the central reconnection region (middle panel), where the magnetic field clearly reverses its direction on the mid plane. The current density in the sheet exceeds the threshold for dissipation into ohmic heating imposed in the simulation ($J_{\mathrm{cr}} = 250\,\mathrm{Fr}\,\mathrm{cm}^{-2}\,\mathrm{s}^{-1}$).
The current density is rather weak at the beginning; it intensifies first in the $D_{\mathrm{R}}$, and then in the region around the collimated jets, where it fragments.
%A switch in the orientation of the \paolo{horizontal component of the magnetic field} occurs across a narrow region close to the centre of the box i.e. where the reconnection jet is accelerated, as shown by the black arrows, representing the projection of the magnetic vector field on the plane. In particular, it abruptly diminishes and eventually cancels where a strong current sheet develops and eventually exceeds the threshold value of dissipation.
%In the \paolo{third} column of Fig. \ref{Fig:dynamics_nanojet} 
The dissipation of the reconnection current sheets into heat is confirmed in the temperature maps, with very high values ($T \gtrsim 8\,\mathrm{MK}$), although rapidly decreasing 
%finally displays the temperature distribution. In particular, \paolo{the ohmic dissipation of strong currents at reconnection region leads to high-temperature spots  ($T \gtrsim 8\,\mathrm{MK}$) that quickly vanish 
because of the effect of expansion and thermal conduction. The reference outflow itself is actually at these high temperatures while the density does not exceed $n \sim 10^9\, \mathrm{cm}^{-3}$  (see appendix \ref{Sec:Athmospheric_stratification} for more details).
%as the strongest current deposits around the reconnection centre, high-temperature spots  ($T \gtrsim 8\,\mathrm{MK}$) reside within these dissipation regions, indicating that ohmic dissipation likely plays a role in heating the surrounding plasma.

%\section{Nanojet diagnostics}
%\label{Sec:Nanojetdiagnostics}
%So far, we have described the dynamics of a nanojet event occurring in a tenuous atmosphere heated up to $10\,\mathrm{MK}$. 
%In this section we discuss the result concerning the synthetic plasma diagnostics we extracted from the simulation results, described so far. 
%In order to test the possibility to detect such reconnection events in the solar corona, we synthesise the emission lines of highly ionised iron atoms which have a formation temperature close to the one found during the nanojet reconnnection in the MHD simulation.

The evolution of the reconnection outflow in the midst of the first $300\,\mathrm{s}$ from the onset of the instability is shown in Fig. \ref{Fig:nanojet_time_evolution}. 
The top panel shows the X-component of the velocity averaged in a box of size $\Delta x = 10\,\mathrm{Mm}$, $\Delta y = 4 \,\mathrm{Mm}$, $\Delta z = 10\,\mathrm{Mm}$, centred at the origin, as a function of time and $x$.
%averaging along a narrow slit $\Delta y = 4\,\mathrm{Mm}$ wide.
The plasma is expelled in opposite directions  at $t \sim 170 s$. 
At the same time, the current density and the temperature are  already high ($J > J_{cr}$, second panel, $T \sim 6\,\mathrm{MK}$, third panel), and grow even higher ($J \sim 400$, $T \sim 8\,\mathrm{MK}$) in about 10 s.
%Then, secondary current sheets are advected by reconnected field lines, as they expand trought the background field. 
%and the temperature shifts to $8\,\mathrm{MK}$ (third panel). 
%Then, as the high temperature plasma expands outward, it is eventually further heated by compression, (numerical) viscosity, and secondary reconnection episodes.
The velocity stays high ($\gtrsim 200\,\mathrm{km}\,\mathrm{s}^{-1}$) for about $30\,\mathrm{s}$, then the jet slows down, the current density dissipates, and the plasma smoothly cools down. We can therefore estimate as $\sim 30\,\mathrm{s}$ the outflow overall duration.
The two opposite jets propagate (toward positive and negative $\hat x$, respectively) with different velocities (as also remarked by the different front slopes in the first panel of Fig. \ref{Fig:nanojet_time_evolution}). In particular, the bidirectional jet, after $\Delta t = 30\,\mathrm{s}$ the reconnection takes place, has expanded by about $10\,\mathrm{Mm}$, with $\sim 40\%$ of asymmetry between the two parts.

During the event, a magnetic energy amount of $\sim 10^{24}\,\mathrm{erg}$ is converted into kinetic and internal energy, as shown in the bottom panel of Figure \ref{Fig:nanojet_time_evolution}, which shows the evolution of the total magnetic, kinetic, and internal energy over the entire elapsed time of the MHD avalanche in the slab where the outflow dynamics is stronger. This energy budget is compatible with the nanoflare energy predicted by \cite{parker1988nanoflares}.
%, and agrees with the observational and numerical results discussed by \citep{antolin2021reconnection}. 
Within the box, an excess of magnetic energy initially increases, but then rapidly drops (below its initial level) upon activation of the anomalous resistivity. Concurrently, both the thermal and kinetic energies rise at similar rates. Magnetic and thermal energy variations are larger compared to kinetic variations: plasma compression concurs in heating at the expense of the kinetic energy, localized near the centre of the reconnection (where the plasma is accelerated to $\sim200\,\mathrm{km}\,\mathrm{s}^{-1}$).
The peak in kinetic energy lasts about $30\,\mathrm{s}$, compatibly with the estimated duration of the outflow.

\section{Forward modelling and diagnostics}
Having described the physical processes that lead to a reconnection outflow in our MHD avalanche simulation, we then test the detection of such an event. Specifically, the question is whether the outflow can be detected in the EUV band. %that contain spectral lines emitted by such hot plasma. 
%According to the temperatures involved in Fig. \ref{Fig:dynamics_nanojet} and \ref{Fig:nanojet_time_evolution}, 
We considered the $94\,\AA$ channel of the Atmospheric Imaging Assembly \citep[AIA,][]{lemen2012atmospheric}  on-board the Solar Dynamics Observatory \citep[SDO,][]{pesnell2012solar}, including the \fexviii\ line emitted at $\log{T} \sim 6.8$, and the $108\,\AA$ \fexix\ spectral line, formed around $\log{T} \sim 7.0$, which will be observed by the forthcoming Multi Slit Solar Explorer (\citealt[MUSE,][]{de2022probing, cheung2022probing}; see Fig. \ref{Fig:Instrument_filters} in the Appendix \ref{sec:appendix_1} which also describes how the synthetic observables are calculated).
Both bands are particularly suitable for detecting the outflow with an estimated apex temperature of $8\,\mathrm{MK}$. Appendix \ref{sec:appendix_1} discusses  more generally the emission in the six EUV AIA channels and in the three MUSE lines (Fig. \ref{Fig:Instrument_filters}).
We have also convolved all the emission maps at the original resolution with the instrumental Point-Spread-Functions (PSFs) and then re-binned them to the instrument pixel size. MUSE PSF is modelled by a Gaussian with FWHM of $0.45"$. AIA PSFs are described in, e.g., \cite{poduval2013point}.

In the synthetic maps, for a more realistic representation, the flux tube box is remapped onto a curved loop-like geometry
(as in \citealt{cozzo2024coronal}).
%is remapped onto a 3D cartesian domain extending over the solar surface in which the loop is represented in a more realistic arc-like structure  
%Such remapping is key for the synthesis of observations with realistic line of sights.
igure \ref{Fig:AIA_diagnostics} presents how we would detect the emission in the AIA $94\,\AA$ channel integrated along a line of sight from a side view of the loops. 
The model has been remapped and oriented along the selected line-of-sight to maximize the brightness of observational signatures at the apex. The overlapping magnetic field lines at the loop top align the hot plasma along the line-of-sight within a compact region, thereby increasing the emission filling factor.

Under nominal operations, AIA exposure times are up to $2.9\,\mathrm{s}$ 
while the basic time step between snapshots is set to $12\,\mathrm{s}$  \citep{lemen2012atmospheric}. We assume an exposure time of $\sim 9\,\mathrm{s}$, to sample an event $\lesssim 36\,\mathrm{s}$ long ($3 \times 12\,\mathrm{s}$ merged observing windows) from $t = 0\,\mathrm{s}$. On the left, we show the emission map in the whole filter band. The low dense and cooler regions of the image are very bright because the filter band includes other intense  $1\,\mathrm{MK}$ lines \citep{testa2012hinode, boerner2014photometric}.
%derived from the AIA \fexviii\ filter. The loop structure is observed from an limb-off and edge-on perspective with 
%Notably, the temperature response function of the AIA filter reveals an additional peak at lower temperatures ($\leq 1\,\mathrm{MK}$) alongside the dominant high-temperature line. 
%This characteristic poses a challenge for nanojet detection since a significant portion of the emission comes from the lower solar corona's cool plasma, known as `coronal moss'\citep{berger1999moss, warren2008observation, testa2013observing} [forse qua ci vogliono altre references].
%Although the moss temperature is far from the main peak of the AIA filter, its emission measure can be significant due to the relatively high plasma density expected in the lower corona.

The hot outflow emission is already visible, but very faint, high ($z \sim 15\,\mathrm{Mm}$) in the image. 
To enhance its contrast, we subtracted the cooler component (l of Fig. \ref{Fig:AIA_diagnostics}) as obtained from other properly rescaled AIA channels (\citealt{reale2011solar, warren2012systematic, cadavid2014heating, antolin2024decomposing}, see also Appendix \ref{sec:appendix_1}). The result is shown in the right panel. The signature of the jet is the horizontal elongated feature high in the image with a bright spot on the left. The brightest emission (above $50\%$ of the peak) is about 10 Mm long.
In this synthetic image, the emission from the hot outflow plasma leads to counts about 4 times higher than the rest of the image, but nevertheless barely detectable without rebinning.
%is presented in the rightmost panel of Fig. \ref{Fig:AIA_diagnostics}, where the nanojet contrast is distinctly amplified, revealing a discernible double-jet-like structure within the high-temperature outflow.
%However, the paucity of photon counts ($\leq 4 \,\mathrm{DN} \,\mathrm{pix}^{-1}$) resulting from the synthesis makes such event challenging to detect in the solar corona using AIA.

In Fig. \ref{Fig:MUSE_diagnostics}, we present  butcorresponding synthetic emissionthe MUSE \fexix\ 108 \AA\ spectral line. We considered an observing mode with a long exposure time of $30\,\mathrm{s}$ from $t = 0\,\mathrm{s}$.
This exposure time envelopes the event completely, although the bulk of the emission is contained in a shorter time, as shown in Fig. \ref{Fig:nanojet_time_evolution} and new movies A1 and A2 in Appendix \ref{sec:appendix_1}.
The line of sight is the same as in Fig. \ref{Fig:AIA_diagnostics}. A 3D rendering of the \fexix\ line emission is shown in Fig. \ref{Fig:Nanojet_3D_side} (Appendix \ref{sec:appendix_1}).
%In the first column, the loop structure is depicted from a `face on' perspective, revealing the arc-like structure of the flaring plasma. The upper panel exhibits the intensity of \fexix\ emission line, highlighting the region atop the structure, where the nanojet forms, as the site of highest emission. The middle panel showcases Doppler shifts, clearly showing the nanojet forming at the interface between blue-shifted and red-shifted regions located near the loop apex at heights of $z \approx 1.5 \times 10^9\,\mathrm{cm}$. These regions, characterized by oppositely directed bulk motions, are correlated with the swirling plasma flows discussed in section \ref{Sec:Nanojetdynamics}. They move in opposite directions mainly along the line of sight and drive the merging of misaligned magnetic field lines at the location of the nanojet acceleration. The lower panel on the left shows the non-thermal line broadening. The strongest widths are observed at nanojet locations, as well as between the misaligned bundles of field lines involved in the reconnection process.
%In the second column of Fig. \ref{Fig:MUSE_diagnostics},at the top, 
To improve for photon statistics closer to the detection level \citep{de2020multi}, we rebin the map on macro-pixels ($0.4" \times 2.7"$).
The \fexix\ emission map is very similar to the "hot" 94 \AA\ map on the right of Fig.\ref{Fig:AIA_diagnostics}, and it highlights the  bipolar jet about 500\% more clearly, and the hot emission is better isolated in this single line.
Also in this case the brightest emission (above $50\%$ of the peak) has an elongated shape, about $10\,\mathrm{Mm}$ long.
%, $\sim 10^9\,\mathrm{cm}$ long, $\sim 0.3 \times 10^9\,\mathrm{cm}$ wide and inclined by $\sim 10$ degrees. 
%,  to increase the level of emission closer to the detection level \citep{de2020multi}. 
%In this case, the emission comes mostly from the region around the loop apex, where the temperature reaches $\sim\,10\,\mathrm{MK}$.
%In this case, the photon counts are higher (approximately $\,\mathrm{DN}\,\mathrm{pix}^{-1}$), rendering this event detectable by the MUSE spectrometer within an exposure time of $10\,\mathrm{s}$ \citep{de2020multi}.
The \fexix\ line profile (on the right) is obtained by integrating over an area $5 \times 5$ Mm$^2$ large and inclined by $10$ degrees (schematically highlighted on the right, and between the white lines of the mid panel) and using a spectral bin of $80\,\mathrm{km}\,\mathrm{s}^{-1}$ \citep[twice the selected for MUSE][]{de2020multi}. The surface is oriented exactly along the jet, and in this way we emphasize the plasma motion along the line of sight.  The double-sided jet determines the presence of a clearly defined double peak in the line profile, with peaks located at $v \approx \pm 200\,\mathrm{km}\,\mathrm{s}^{-1}$ (the Alfvén velocity is about $1000\,\mathrm{km}\,\mathrm{s}^{-1}$). %\delete{Non-thermal line broadening shows the same order of magnitude when measured from a "side-on" perspective (i.e. with the jets oriented towards/away from the observer) and close to reconnection site (see Fig. \ref{Fig:XZ_FM} for more details).}

%Finally, the last panel portrays a scatter plot of the velocity component aligned with $\vec B$ as a function of temperature, sampled pixel by pixel within the same reference box as before. Each point is colour-coded according to its emission in the \fexix\ line. The point distribution illustrates a clear trend below $\sim 6\,\mathrm{MK}$, where plasma velocity proportionally increases with temperature. In particular, above $\sim 5\,\mathrm{MK}$, both velocity and plasma emission show rapid escalation, suggesting their association with the accelerated and heated nanojet. The hottest points ($T > 6\,\mathrm{MK}$) exhibit slower velocities, as they are located near the reconnection site where the tension force has not yet accelerated the plasma outflow.

\section{Discussion and conclusions}
\label{Sec:Conclusions}
%This study delves into the dynamics and diagnostics of nanojets arising during a magnetohydrodynamic (MHD) avalanche process within the solar corona. Specifically, 
In this work we study the serendipitous formation and evolution of a reconnection outflow within the complex background of a coronal flux tube system, which fragments into smaller current sheets with random reconnection episodes. We analyse what kind of detection we might expect both with current instruments, such as the AIA imager, and with the forthcoming MUSE spectrometer.This outflow is the result of a reconnection event, of nanoflare size, and comes out perpendicular to the flux tube guide field. It shares therefore many features with observed small size jets, named nanojets \citep{antolin2021reconnection}.

Previous 3D MHD simulations had addressed the issue of nanojets acceleration with ad hoc setups where field lines are tilted  by photospheric motions \citep{antolin2021reconnection} or where their misalignment is provided from the outset, by initial conditions \citep{pagano2021modelling}.
%\sout{They provide a clear description of nanojet acceleration in terms of occurrence of magnetic reconnection and release of magnetic field tension perpendicularly to the guide magnetic field.}

\cite{cozzo2023coronal} 3D MHD model describes the turbulent, large-scale energy release of multiple magnetic strands within a stratified atmosphere, twisted by footpoint motions.
This model provides an excellent opportunity to study the development of jets and possibly nanojets, and their possible detection, where they are dispersed in a more realistic situation. 
rk we single, out a magnetic reconnection event based on the heating and plasma acceleration that it causes, in the midst of the dynamic and thermally evolving loop structures.

The described event is localized. It involves the thick field lines in Fig. \ref{Fig:Nanojets_3D}, which are driven to cross each other and then detach again with a different topology. Although the configuration is not ideal as in plane parallel cases, there are all signatures of reconnection, including localized heating and perpendicular flows. The $E_{\parallel}$ component can be different from zero only in non-ideal plasma conditions (reconnection) and shows very high values halfway down the loop, where the field lines initially cross each other, and much smaller nearby, as emphasized in Fig.\ref{Fig:Nanojets_3D_A} showing the extension of the dissipation region ($E_{\parallel} \ne 0$) in full 3D.
Similarly to \cite{antolin2021reconnection} observations, the jet is observed after the initial MHD avalanche.
The outflow event is generated as a result of the formation and dissipation of a current sheet, induced by the chaotic motion of plasma and magnetic field lines during the MHD avalanche. 
The evolution of the magnetic field lines is remarkably similar to the schematic picture shown in Fig. \ref{Fig:reconnection}.
This event exhibits typical signatures of nanojets as anticipated by previous theoretical and numerical investigations, and observations \citep[e.g.,][]{antolin2021reconnection, sukarmadji2022observations}, such as typical lateral dimensions of a few thousand kilometres and typical velocity of a few hundreds of kilometres per second. It also takes place at the top of the loop, as many nanojets observed by \cite{antolin2021reconnection, sukarmadji2022observations, sukarmadji2024transverse}. The magnetic energy released during this event approximates $10^{24} \,\mathrm{erg}$, in agreement with \cite{parker1988nanoflares}. A significant fraction of this energy is converted into heat, increasing the plasma temperature above $8 \,\mathrm{MK}$, while the remaining energy propels the outflow.
The jets originate from a localized reconnection event that is the dominant source of plasma heating. Field lines overlapping, when the system is mapped into a curved geometry, increases the plasma filling factor and, therefore, the signatures in the forward modelling.
ref{Fi5that the outflow described here is very difficult to detect with present-day capabilities (AIA).
Specifically, the AIA $94\,\mathrm{\AA}$ channel is in principle sensitive to such hot plasma, but its detection is made difficult both by the low emission measure of these events and by the presence of a strong cool component in the same filter band. The subtraction of this cool component is an approximation that does not work perfectly and introduces an extra source of noise (on top of photon noise, readout noise, digitization noise). Such noise related to the subtraction is likely significantly larger than the other sources of noise and not properly quantifiable because the individual contributions of the different spectral lines within the broad AIA passbands cannot be accurately determined.

It is arguable that a stronger magnetic field, or denser loops, can lead to events easier to detect. In fact, higher heat capacity  $c$ (due to high averaged density) needs higher magnetic energy budgets $\Delta E_B$ to keep the temperature high, i.e., in the \fexviii -\fexix\ temperature range, according to:
\begin{equation}
c\,n\,\Delta T = \Delta \left(\frac{B^2}{8\pi}\right),    
\end{equation}
indicating that $\Delta T\propto B^2/n$.
%As the continuous driver pumped energy balances ultimately radiative losses in the corona, coronal density scales with the overall nanoflare heating.
With such a scaling pattern, the dissipation of a magnetic field just $\sqrt{10}$ times stronger (e.g. $\sim 30\,\mathrm{G}$) heats up a $10$ times denser plasma (e.g. $\sim 10^{10}\,\mathrm{cm}^{-3}$) to million degrees (up to $\sim 10\,\mathrm{MK}$), attaining a $10^2$ larger emission measure (enough to be easily detected by MUSE at a cadence as short as 10 s).
In our simulation, we considered a typical coronal magnetic field strength of $10\,\mathrm{G}$ \citep{long2017measuring} with plasma density of about $10^9\,\mathrm{cm}^{-3}$. Nevertheless, in active regions, magnetic field can exceed $30\,\mathrm{G}$ \citep[e.g.][]{van2008coronal, jess2016solar, brooks2021measurements} while density can reach $10^{10}\,\mathrm{cm}^{-3}$ \citep{reale2014coronal}. In these cases, higher emission is expected and, comparably, shorter exposure times would be needed to single out the outflow jet, making its evolution suitable to be inferred with short-cadence ($10\,\mathrm{s}$ or less) observing modes (that will be available with MUSE).
This scaling needs to be verified to pave the way for the detection of magnetic reconnection in the solar corona. 
he event described has all the physical features predicted by the theory of reconnection and shows strong similarities with
the theoretical model proposed by \cite{antolin2021reconnection}:
they both originate from a small-angle reconnection event, and share the same orders of magnitude in terms of dimensions, velocity, and duration of the outflow jets. A detailed, physical analysis of the simulated event has also shown many features matching with \cite{antolin2021reconnection} numerical model, including: the distribution and orientation of the velocity field (Fig. \ref{Fig:Nanojets_3D}), the detailed evolution of the magnetic, kinetic, and thermal energy (Fig. \ref{Fig:nanojet_time_evolution}), and the location and structuring of the dissipation region (Fig. \ref{Fig:Nanojets_3D_A}). As a significant deviation from the more idealized model by \cite{antolin2021reconnection}, we have shown a bidirectional, but asymmetric jet. Although we do not account for magnetic curvature \citep{pagano2021modelling} in the simulation, other factors, in particular local field line braiding and warping, and the non-uniform background plasma effectively make the propagation different on the two sides.

The \cite{antolin2021reconnection} model suggests the role of small-angle magnetic reconnection in accelerating (nano-) jets within a non-vanishing coronal loop magnetic field.
This is supported by observational evidence of collimated jets, interpreted as the kinetic counterpart of nanoflare heating.
Smoking guns of such nanoflare heating in the tenuous solar corona are difficult to catch because of the small emission measure, and the highly efficient thermal conduction, limiting the visibility of such events to their already short kinetic time scales.
In this work we show that detection of nanoflare jets might be possible with MUSE, even at high temperatures, when the plasma is  under-dense and fainter, thanks to 
%. In particular, MUSE will allow for 
the MUSE detailed EUV spectroscopic diagnostics, until now restricted to the UV band \citep{de2014interface}.

The onset of the reconnection event is caused by the overlapping of two misaligned bundles of field lines, ultimately brought together by the residual dynamics of the MHD avalanche. This scenario supports the interpretation of \cite{antolin2021reconnection}, and \cite{sukarmadji2022observations} observations, where it is argued that MHD instabilities can trigger reconnection nanojets.

occurring  around the MHD avalanche triggered by the kink instability. At that time, the impulsive heating events have not been effective in filling the flux tubes with dense plasma yet. In these conditions of tenuous plasma, the heat pulse is effective in determining a steep increase of the local temperature, and the outflowing jet is therefore hot as well, and faint because of the low density. 

%increases  The temperature of the nanojet is determined by the local  conditions. The nanojet is so hot because it is triggered just after the impulsive phase of the kink instability.This transient phase marks the initial brightening of the loop, when the flux tube is still relatively empty of plasma. As densities involved are still relatively low, the energy released by the reconnection determines a large temperature increase. This makes the local heating particularly effective and is ultimately the main reason why such hot plasma is involved, which may make the detection easier, but at low emission levels. 

This period of the evolution probably represents a relatively short transient in the global evolution of a loop system. 
So such hot and faint jets are also probably infrequent and fainter than nanojets observed so far \citep[e.g.][]{antolin2021reconnection}. In future work, we intend to study the formation and possible detection of reconnection jets in more steady-state conditions, i.e., later in the loop evolution, when the flux tubes are filled with denser plasma coming up from the chromosphere, driven by the heating. Cooler and brighter nanojets are therefore expected later in this more steady-state regime.

\begin{acknowledgements}
      GC, PP, and FR acknowledge support from ASI/INAF agreement n. 2022-29-HH.0. 
      This work made use of the HPC system MEUSA, part of the Sistema Computazionale per l'Astrofisica Numerica (SCAN) of INAF-Osservatorio Astronomico di Palermo.
      %JR and AWH acknowledge the financial support of Science and Technology Facilities Council through Consolidated Grant ST/W001195/1 to the University of St Andrews.
      PT was supported by contract 4105785828 (MUSE) to the Smithsonian Astrophysical Observatory, and by NASA grant 80NSSC20K1272x.
      BDP and JMS were supported by NASA contract 80GSFC21C0011  (MUSE).
\end{acknowledgements}

\bibliographystyle{aa} % style aa.bst
\bibliography{bibliography}

\begin{appendix} %First appendix

\section{details on forward modelling}
\label{sec:appendix_1}

\begin{figure}[!b]
   \centering
\includegraphics[width=\hsize]{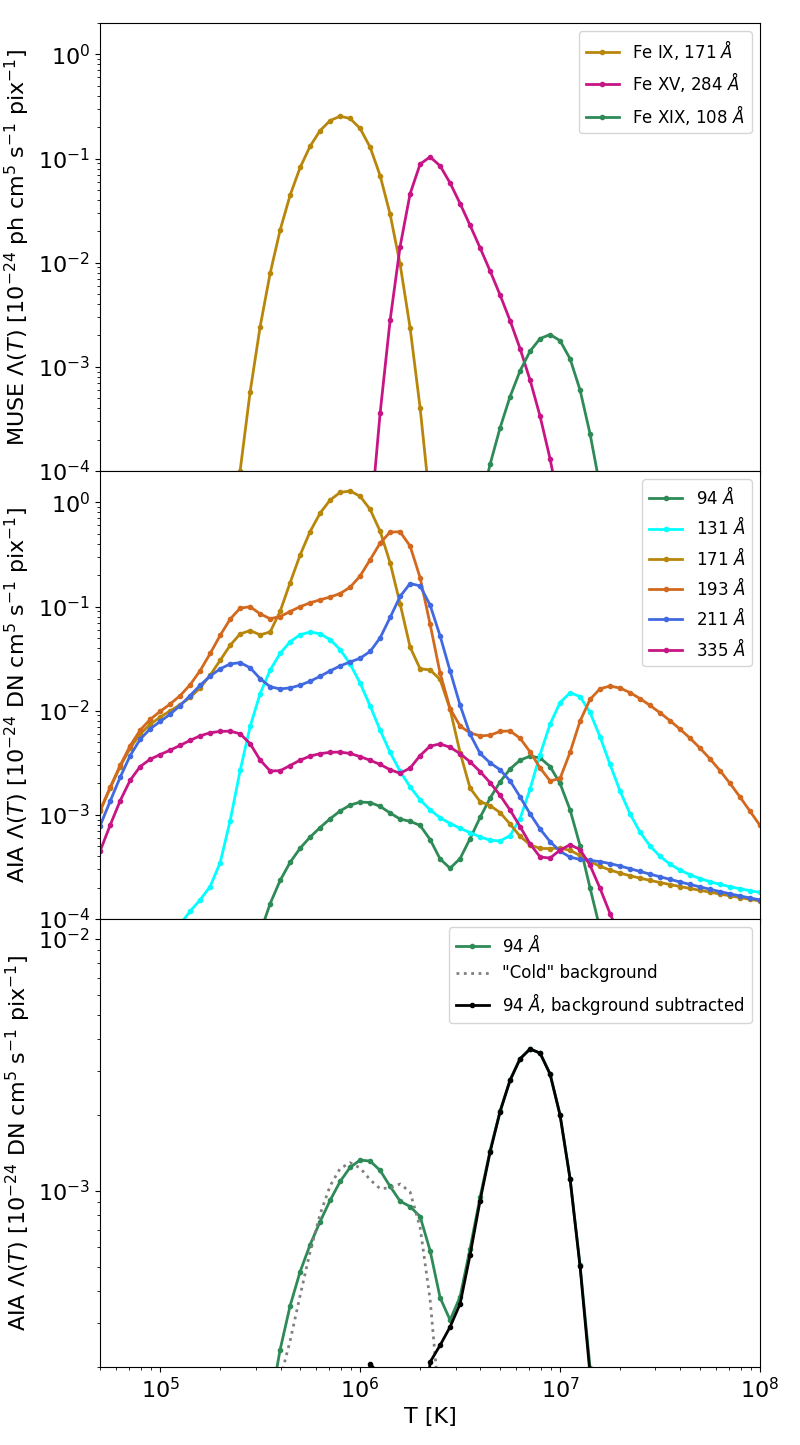}
  \caption{Temperature response functions $\Lambda_i (T)$ for the AIA and MUSE channels. Top: MUSE spectrometer lines: \feix\ $171\,\AA$, \fexv\ $284\,\AA$, and \fexix\ $108\,\AA$. Middle: AIA $94\,\AA$ (containing \fexviii\ line), $131\,\AA$, $171\,\AA$, $193\,\AA$, $211\,\AA$, and $335\,\AA$ channels. Bottom: expected response function of the $94\,\AA$ AIA channel (black line) after the subtraction of the cold component (dotted line) obtained by combination of  $131\,\AA$, $171\,\AA$, $193\,\AA$, $211\,\AA$, and $335\,\AA$ AIA channels from the the total one(light green).
  }
  \label{Fig:Instrument_filters}
\end{figure}

\begin{figure*}[h!]
   \centering
\includegraphics[width=0.95\hsize]{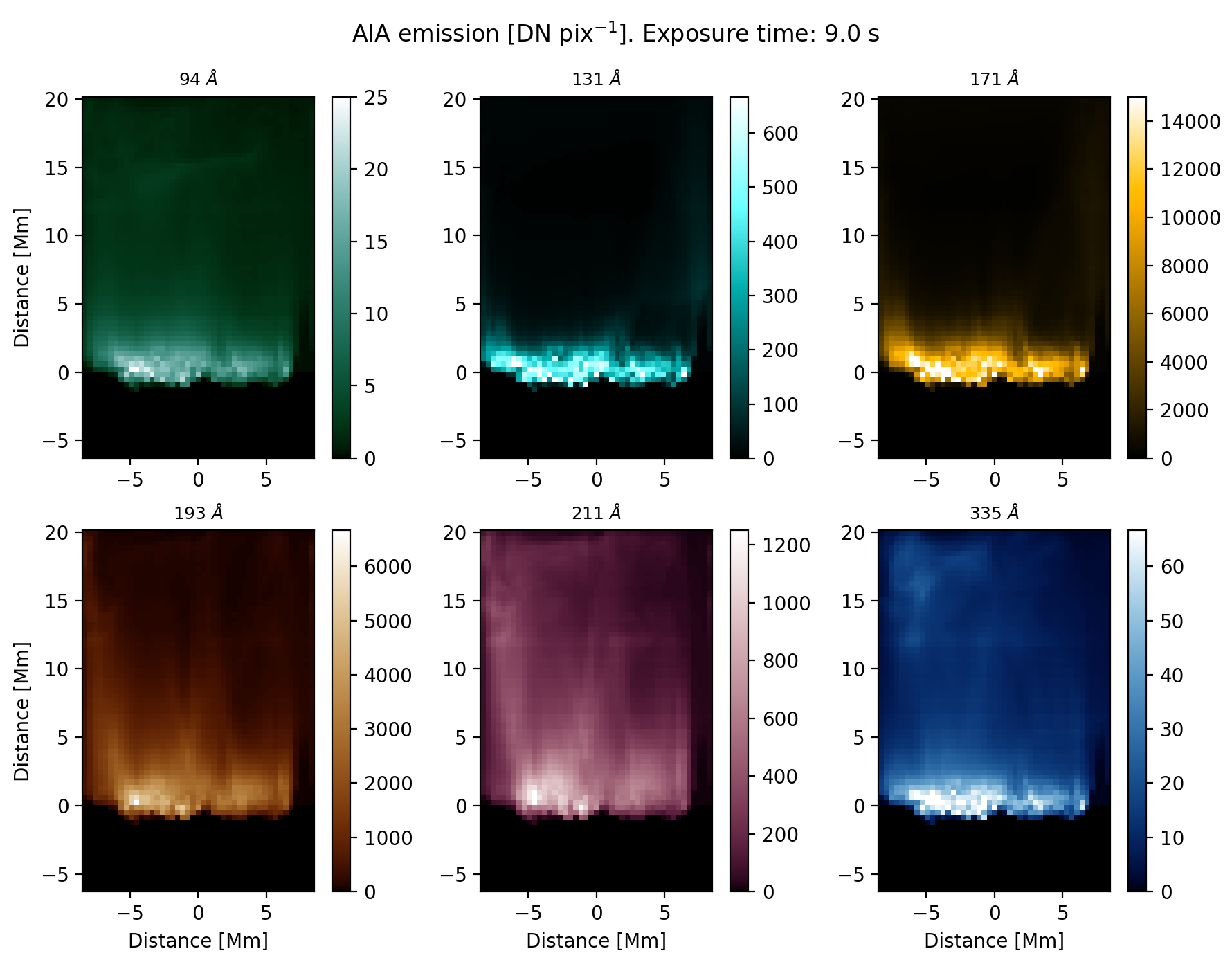}
  \caption{Synthetic maps of AIA emission integrated along a line of sight from a side view of the curved loop system for $94\,\AA$, $131\,\AA$, $171\,\AA$, $193\,\AA$, $211\,\AA$, and $335\,\AA$ channels, respectively (see Movie A1).}
  \label{Fig:AIA_emission}
\end{figure*}

\begin{figure*}[h!]
   \centering
\includegraphics[width=0.95\hsize]{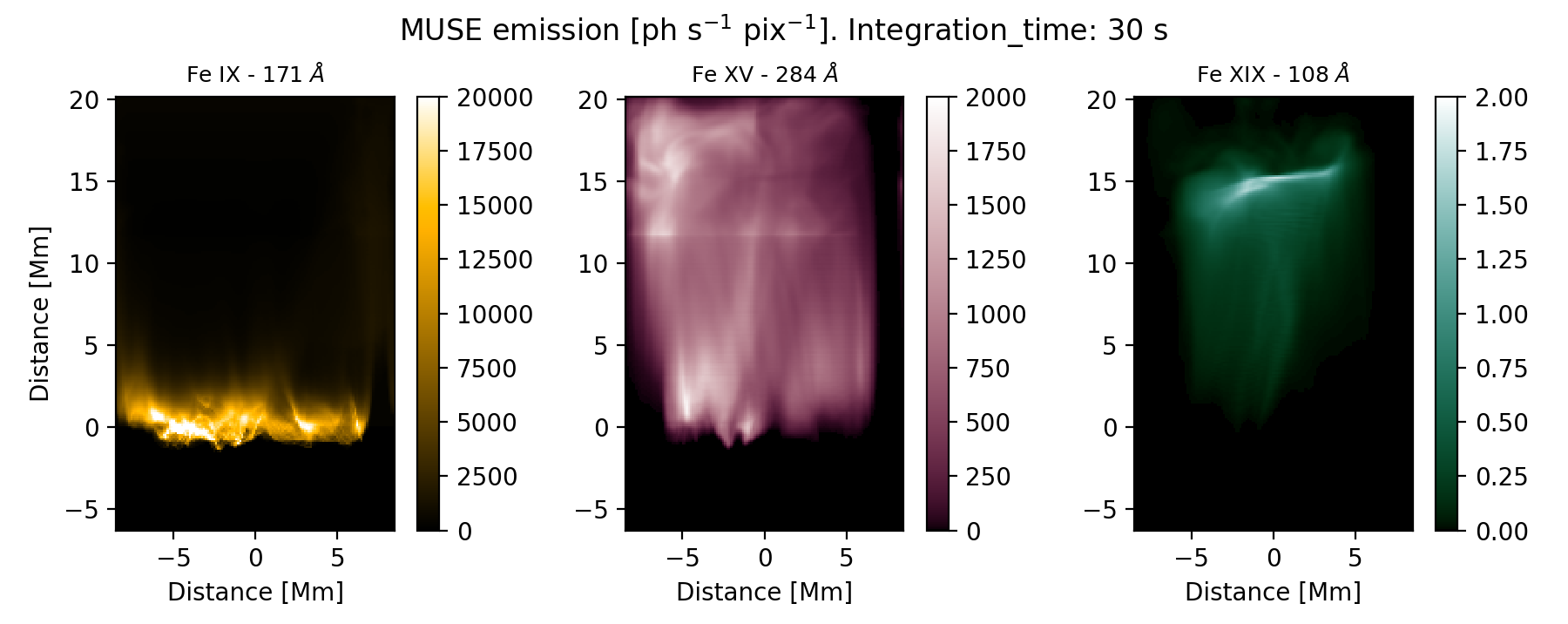}
  \caption{Synthetic maps of MUSE emission  integrated along a line of sight from a side view of the curved loop system for  the \feix\ $171\,\AA$, \fexv\ $284\,\AA$, and \fexix\ $108\,\AA$ lines, respectively (see Movie A2).}
  \label{Fig:MUSE_emission}
\end{figure*}

%\begin{figure}[h!]
%   \centering
%\includegraphics[width=\hsize]{Images/AIA_MUSE.png}
%  \caption{Temperature response function $\Lambda_i (T)$ for the AIA channel at $94\,\AA$ (black curve) and the MUSE emission line of \fexix\ at $108\,\AA$ (orange curve). The blue curve represents the expected response function of the AIA channel after the subtraction of a cold background (dashed line) obtained by combination of AIA channels at $131\,\AA$, $171\,\AA$, $193\,\AA$, $211\,\AA$, and $335\,\AA$.}
%  \label{Fig:AIA_94_filter}
%\end{figure}
%\begin{figure}[h!]
%   \centering \includegraphics[width=\hsize]{Images/XZ_FM.png}
%    \caption{MUSE synthetic observables with the loop in off-limb configuration. The emission is integrated over macro-pixels of size $\Delta_h = 0.28$ Mm, $\Delta_v = 1.89$ Mm ($0.4" \times 2.7"$) (exposure time: $30\,\mathrm{s}$). \textbf{Top panel:}  the \fexix\ 108 \AA\ line emission map. The emission is integrated over macro-pixels as in Fig. \ref{Fig:MUSE_diagnostics}. \textbf{Mid panel:} \fexix\ Doppler shifts map. \textbf{Lower panel}: non-thermal line broadening.}
%    \label{Fig:XZ_FM}
%\end{figure}

\begin{figure}[h!]
   \centering \includegraphics[width=\hsize]{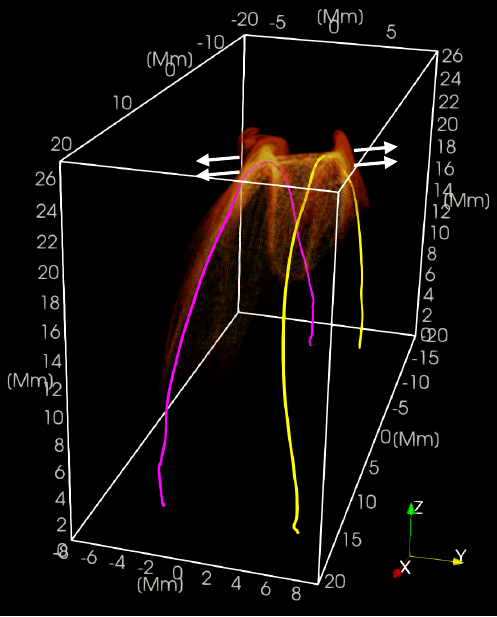}
    \caption{3D rendering of the MUSE \fexix\ line emission at $t=10\,\mathrm{s}$. In the 3D box, we show two field lines remapped in curved geometry for reference. The arrow point in the direction of the jet propagation.}
    \label{Fig:Nanojet_3D_side}
\end{figure}

\begin{figure}[h!]
   \centering
    \includegraphics[width=\hsize]{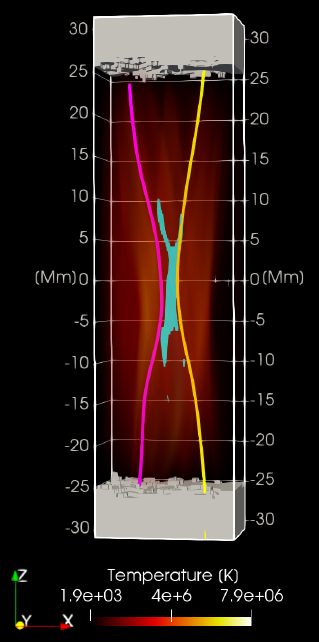}
    \caption{3D rendering of the analysed magnetic reconnection region. Field lines are the same as in the second panel of Fig. \ref{Fig:Nanojets_3D} ($t = 10\,\mathrm{s}$). The temperature (red colour) is shown in the background. Here we also show the diffusion region $D_R$ (pale blue, Fig. \ref{Fig:reconnection}) defined by the region where where $\vec E \cdot \vec B \ne 0$. The solid blocks at the top and bottom are the footpoints in the chromosphere.}
    \label{Fig:Nanojets_3D_A}
\end{figure}

\begin{figure}[h!]
   \centering \includegraphics[width=\hsize]{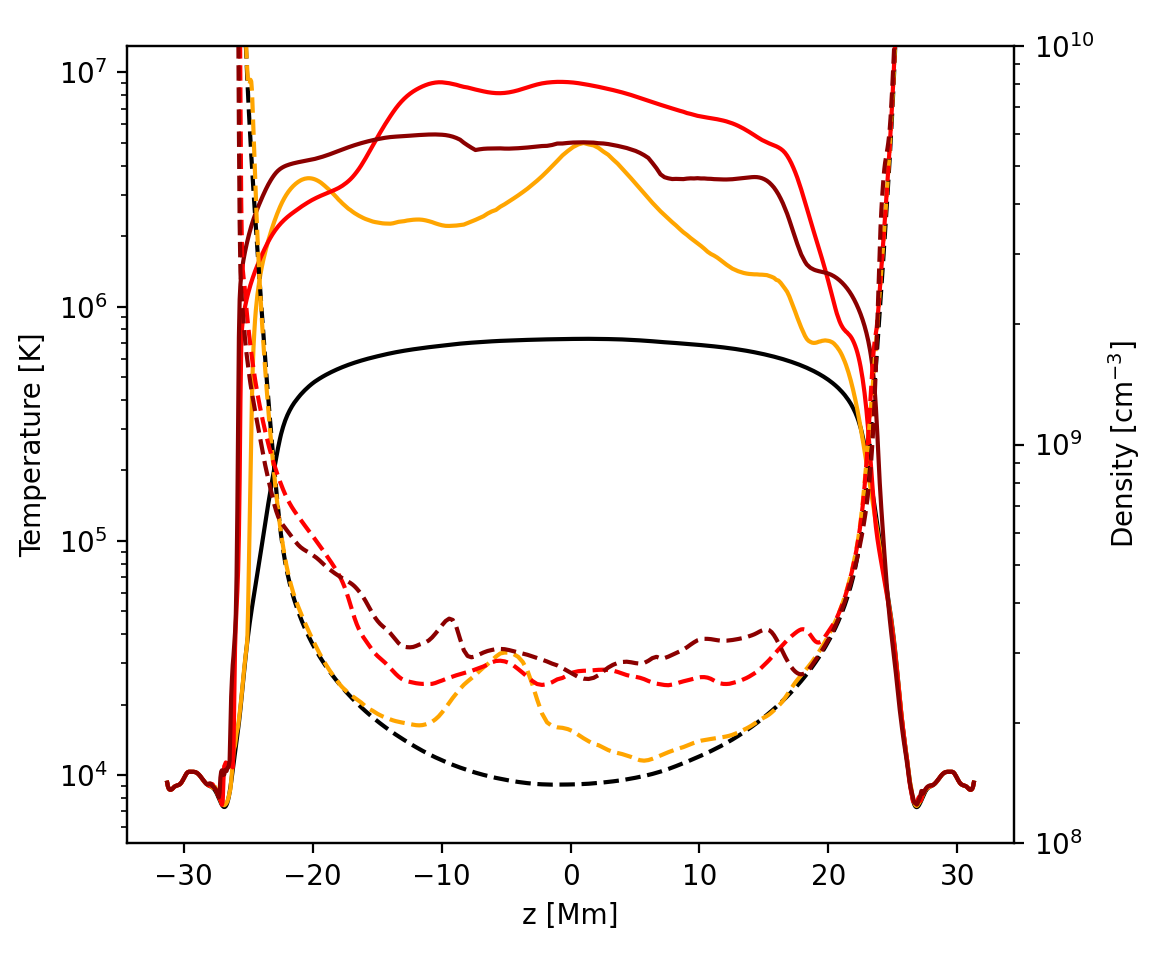}
    \caption{Plasma vertical stratification: temperature (solid line) and density (dashed) along a 1-pixel column crossing the jet centre (coordinates: $x = y = 0\,\mathrm{Mm}$) before the instability (black line) and at times $t = 0\,\mathrm{s}$ (orange) $\Delta t = 10\,\mathrm{s}$ (red), $t = 20\,\mathrm{s}$ (dark-red) from the ignition of the jet.}
    \label{Fig:1D_plot}
\end{figure}
 to compare with the familiar semicircular shape of coronal loops, the original simulation outputs were remapped and interpolated onto a new cartesian grid. In this process, the initially straight coronal section of the domain was curved into a half-cylinder shell with a characteristic radius of $R = L/\pi$ with $L = 50\,\mathrm{Mm}$, the loop's coronal length. On the other hand, the chromospheric layers were modelled as two parallel parallelepipeds, as described in \cite{cozzo2024coronal}.

To remap on observation-like images, we computed line emission $I_{0}$ from the pixel $i,j$, by integrating cells intensity $F_{i,j,k}$ (in units of $\mathrm{ph}\,\mathrm{s}^{-1}\,\mathrm{pix}^{-1}$) along the line of sight \citep[as in ][]{de2022probing, cozzo2024coronal}:
\begin{equation}
I_{0}^{i,j} = \sum_k F_{i,j,k}, \\
\end{equation}
with:
\begin{equation}
    F_{i,j,k} = n_e^2(\tilde x, \tilde y, \tilde z; T) \, \Lambda_f(T) \, \Delta z, 
\end{equation}
where $n_e$ is the free electron density, $\Lambda (T)$ is the instrument temperature response function, and $\Delta z$ is the cell width. Instrument temperature response functions are calculated using CHIANTI 10 \citep{del2021chianti} with the CHIANTI ionization equilibrium, coronal element abundances \citep{feldman1992elemental}, assuming a constant electron density of $10^9\,\mathrm{cm}^{-3}$, and no absorption considered.

%In view of analysing line profiles, Doppler-shifts and non-thermal line broadening are related to the first, and second momenta of the velocity distribution:
%\begin{equation}
%I_1^{i,j} = \frac{\sum_{k} F_{i,j,k} v_{i,j,k}}{I_0},
%\end{equation}
%\begin{equation}
%I_2^{i,j} = \sqrt{\frac{\sum_{k} F_{i,j,k} (v_{i,j,k} - I_{1i,j})^2}{I_0}},
%\end{equation}
%being $v_{i,j,k}$ the velocity in the pixel aligned to the line of sight \citep[as in ][]{de2022probing, cozzo2024coronal}.

%Doppler-shifts and non-thermal line broadening are related to the first, and second momenta of the velocity distribution:
%\begin{equation}
%I_1^{i,j} = \frac{\sum_{k} F_{i,j,k} v_{i,j,k}}{I_0}
%\end{equation}
%\begin{equation}
%I_2^{i,j} = \sqrt{\frac{\sum_{k} F_{i,j,k} (v_{i,j,k} - I_{1i,j})^2}{I_0}},
%\end{equation}
%where $v_{i,j,k}$ is the LOS velocity in each pixel.

In Fig. \ref{Fig:Instrument_filters} we show the temperature response functions $\Lambda (T)$ of the three MUSE EUV lines: \feix\ $171\,\AA$, \fexv\ $285\,\AA$, and \fexix\ $108\,\AA$ \citep[top panel, ][]{de2020multi}; and of the six AIA EUV channels at $94\,\AA$, $131\,\AA$, $171\,\AA$, $193\,\AA$, $211\,\AA$, and $335\,\AA$, respectively \citep[mid panel, ][]{boerner2012initial}.

In Fig. \ref{Fig:AIA_emission} we show synthetic AIA emission maps from $9\,\mathrm{s}$ effective exposures. 
%across a $30\,\mathrm{s}$ observing window. 
Each panel shows the side view of the intensity distribution integrated over the entire filter band of the six EUV channels in Fig. \ref{Fig:Instrument_filters}.
Emission in $131\,\AA$, and $171\,\AA$ channels is dominated by a relatively cool ($\lesssim 1\,\mathrm{MK}$) plasma component just above the transition region ($< 5\,\mathrm{Mm}$ in height);  $193\,\AA$, $211\,\AA$, and $335\,\AA$ channels show evidence of warmer plasma ($2-4\,\mathrm{MK}$) at intermediate height ($\gtrsim 5\,\mathrm{Mm}$). A faint feature from hot ($\gtrsim 5\,\mathrm{MK}$) plasma shows up around the loop top ($\sim 15\,\mathrm{Mm}$) in the $94\,\AA$ channel (containing the `hot' \fexviii\ line), although most of the intensity comes from the cooler plasma background.
Movie A1 shows the evolution of the coronal loop emission as imaged by the six AIA channels (Fig \ref{Fig:Instrument_filters}), and contributing to the integrated emission of Fig. \ref{Fig:AIA_emission}. The atmosphere appears roughly steady in all the channels, with some noisy gleaming in the lower, cooler corona, and slow variations in the atmospheric structuring of the warm plasma at intermediate heights. Only the hot plasma jet  at the loop top evolves dynamically,d expandsgtward until its emission vanishes in the background.

The synthetic emission as sampled by the three MUSE channels (Fig. \ref{Fig:Instrument_filters}) is shown in Fig. \ref{Fig:MUSE_emission}. We assumed a $30\,\mathrm{s}$ exposure time and line of sight from a side view of the curved loop. MUSE lines  (\feix\ 171\AA, \fexv\ 284\AA, and \fexix\ 108\AA) detect plasma emitting mostly around $\sim 1\,\mathrm{MK}$, $\sim 2\,\mathrm{MK}$, and $\sim 10\,\mathrm{MK}$ plasma, respectively.
In particular, the \feix\ line is emitted mostly at the loop footpoints; in the \fexv\ line we see the bulk of the loop; the \fexix\ line shows a transient brightening around the loop apex.%(emphasized by the folding of sheet-like structures at the boundary of misaligned flux tubes \citealt{cozzo2024coronal}).
Similarly to Movie A1, Movie A2 shows the evolution of the loop in the MUSE lines when the jet is visible.  No evidence of the jet is found in the cooler MUSE channels. In the hot 108\AA\ line, the bright feature stretches into a strongly elongated structure.

In Fig. \ref{Fig:Nanojet_3D_side}, the 3D rendering in curved geometry clearly shows the jet emission in the MUSE \fexix\ channel at $108\AA$ to be perpendicular to the guide field (represented by the drawn field lines). This can be compared to observed nanojets as in \cite{antolin2021reconnection}.
Specifically, the brighter plasma (yellow volume) envelopes the jet at the loop top, but ``tails'' of hot plasma (redder parts), propagating along the field, also stand out in the EUV \fexix\ line emission. They form because thermal conduction efficiently spreads heat from the reconnection site.

Fig. \ref{Fig:Instrument_filters}
shows also the temperature response function $\Lambda (T)$ for the AIA filter at $94\,\AA$ \citep[light green curve,][]{boerner2012initial} 
%and the MUSE emission line of \fexix\ at $108\,\AA$ \citep[orange curve,][]{de2020multi}. 
The figure also shows the same response function after subtraction of the cool component (dotted line) obtained by a combination of the other AIA responses  ($131\,\AA$, $171\,\AA$, $193\,\AA$, $211\,\AA$, and $335\,\AA$). More in particular, we derived the background-subtracted emission maps $\tilde I_0^{94 \AA}$ for the AIA filter at $94\,\AA$ as follows:
\begin{align}
    \tilde I_0^{94 \AA} &= I_0^{94 \AA} - I_0^{\mathrm{bkg}} \nonumber \\
    I_0^{\mathrm{bkg}} &=
    (2.3 \, I_0^{131 \AA} + 0.8 \,I_0^{171 \AA} + 1.0 \, I_0^{193 \AA} + \nonumber \\ &+ 2.6 \, I_0^{211 \AA} + 30.1 \, I_0^{335 \AA}) \times 10^{-3} 
\end{align}
where $I_0^{\mathrm{bkg}}$ is the cooler background image we obtain from the other AIA filters \citep{reale2011solar, warren2012systematic, cadavid2014heating, antolin2024decomposing}. In this way we manage to isolate better the emission in the hot \fexviii line.

To compute the line profiles (\fexix\ line), we assumed at fixed temperature, density, and plasma velocity, a Gaussian profile:
\begin{equation}
    f_{\mathrm{cell}} (v) = \frac{F_{\mathrm{cell}}}{\sqrt{2 \pi \sigma_T^2}} \exp{ \left[ -\left(\frac{v - v_{\mathrm{cell}}}{\sigma_T} \right)^2 \right] } 
    \label{Eq:Gaussian_prof}
\end{equation}
where 
$\sigma_T = \sqrt{\frac{2 k_B T_{\mathrm{cell}}}{m_{\mathrm{Fe}}}}$
is the thermal broadening, $m_{\mathrm{Fe}}$ is the Fe atomic mass, and $v_{\mathrm{cell}}$ is the plasma velocity parallel to the line of sight in a single cell. MUSE \fexix\ spectral bin is $\Delta v = 40\,\mathrm{km}\,\mathrm{s}^{-1}$ \citep{de2020multi}. We assumed a spectral bin twice as large, $\Delta v = 80\,\mathrm{km}\,\mathrm{s}^{-1}$, to increase the photon counts. We account for both thermal and non-thermal broadening.
%, but we neglect the instrument broadening.

Finally, we rebinned MUSE \fexix\ observables on macropixels ($0.4" \times 2.7"$) that collect more photon counts, closer to the detection level \citep{de2020multi}; AIA $94\,\AA$ channel intensity is shown with the original pixel size ($0.6" \times 0.6"$).

\section{other 3D renderings}
\label{sec:appendix_1b}

Figure \ref{Fig:Nanojets_3D_A} shows the 3D rendering of the diffusion region (in solid cyan, see also Fig. \ref{Fig:reconnection}) between two reconnecting field lines at $t=10\,\mathrm{s}$ (as in Fig. \ref{Fig:Nanojets_3D}). The thin diffusion region develops where the field lines meet and reconnect (i.e. close to the box centre). It is elongated and oriented along the guide field, with short ``branches'' inclined with the magnetic field bundles to form an ``X shape''.  
Hot coronal plasma is in the proximity of the reconnection site. Field lines are embedded in the chromosphere, shown by solid blocks at the top and bottom sides of the box.
%Fig. \ref{Fig:AIA_diagnostics} and \ref{Fig:MUSE_diagnostics}) shows the curvature of the loop with no overlapping footpoints.
%The upper panel exhibits the intensity of \fexix\ emission line, highlighting the region atop the structure, where the nanojet forms, as the site of highest emission. 
%The middle panel showcases Doppler shifts, with the nanojet forming at the interface between blue-shifted and red-shifted regions, at height of $z \approx 15\,\mathrm{Mm}$, corresponding to oppositely directed bulk motions.
%The lower panel shows non-thermal line broadening. As anticipated in Fig. \ref{Fig:MUSE_diagnostics}, the strongest widths are observed at nanojet locations, as well as along the misaligned bundles of field lines involved in the reconnection process.}

\section{atmospheric stratification}
\label{Sec:Athmospheric_stratification}

The simulated solar atmosphere consists of a chromospheric and a coronal column separated by a thin transition region. Specifically, field-aligned gravity, thermal conduction, optically thin radiative losses, heating by anomalous magnetic resistivity, and background heating structures a $5\,\mathrm{Mm}$ long coronal loop, while its chromospheric footpoints are $\sim 6\,\mathrm{Mm}$ wide each and $10^4\,\mathrm{K}$ hot \citep{reale20163d, cozzo2023coronal}.

Figure \ref{Fig:1D_plot} shows the temperature and density stratification along a column of pixels passing through the jet centre (coordinates: $x = y = 0\,\mathrm{Mm}$). Before the avalanche (black curve), the atmosphere is initially tenuous ($n \sim 10^8\, \mathrm{cm}^{-3}$) and cold ($T \lesssim 1\,\mathrm{MK}$). After the instability,  when the jet is formed, the plasma is rapidly heated (solid, orange line, $t = 0\,\mathrm{s}$) and the temperature rises up to $10\,\mathrm{MK}$ (red line, $t = 10\,\mathrm{s}$), and subsequently cools down (orange line, $t = 20\,\mathrm{s}$). The density (dashed lines) increases as well, approaching, but never exceeding, $n \sim 10^9\, \mathrm{cm}^{-3}$.

\end{appendix}

\end{document}